\documentclass[superscriptaddress,aps,prd,nofootinbib,preprint]{revtex4-1}
\pdfoutput=1

\usepackage{graphicx}
\usepackage[colorlinks,allcolors=blue]{hyperref} 
\usepackage{amsmath,amssymb,bm}
\usepackage[T1]{fontenc}
\usepackage{mathtools}
\usepackage[margin=0.9in]{geometry}

\begin{document}

\title{Misalignment production of vector boson dark matter from axion-SU(2) inflation}

\author{Tomohiro Fujita}
\email{tomofuji@aoni.waseda.jp}
\affiliation{Waseda Institute for Advanced Study, Waseda University, 3-4-1 Okubo, Shinjuku, Tokyo 169-8555, Japan}
\affiliation{Research Center for the Early Universe, The University of Tokyo, Bunkyo, Tokyo 113-0033, Japan}
\author{Kai Murai}
\email{kai.murai.e2@tohoku.ac.jp}
\affiliation{Department of Physics, Tohoku University, Sendai, 980-8578, Japan}
\author{Kazunori Nakayama}
\email{kazunori.nakayama.d3@tohoku.ac.jp}
\affiliation{Department of Physics, Tohoku University, Sendai, 980-8578, Japan}
\affiliation{International Center for Quantum-field Measurement Systems for Studies of the Universe and Particles (QUP), KEK, Tsukuba, 305-0801, Japan}
\affiliation{Kavli IPMU (WPI), UTIAS, The University of Tokyo, Kashiwa, Chiba, 277-8583, Japan}
\author{Wen Yin}
\email{yin.wen.b3@tohoku.ac.jp}
\affiliation{Department of Physics, Tohoku University, Sendai, 980-8578, Japan}

\begin{abstract}
    We present a new mechanism to generate a coherently oscillating dark vector field from axion-SU(2) gauge field dynamics during inflation.
    The SU(2) gauge field acquires a nonzero background sourced by an axion during inflation, and it acquires a mass through spontaneous symmetry breaking after inflation. 
    We find that the coherent oscillation of the dark vector field can account for dark matter in the mass range of $10^{-13}\,\text{--}\,1$\,eV in a minimal setup. In a more involved scenario, the range can be wider down to the fuzzy dark matter region.
    One of the dark vector fields can be identified as the dark photon, in which case this mechanism evades the notorious constraints for isocurvature perturbation, statistical anisotropy, and the absence of ghosts that exist in the usual misalignment production scenarios.  
    Phenomenological implications are discussed.
\end{abstract}

\preprint{TU-1216, KEK-QUP-2023-0033}
\maketitle

\section{Introduction}
\label{sec: intro}

The existence of dark matter is well established by astronomical and cosmological observations.
However, the nature of dark matter remains unknown, and various candidates have been proposed in the literature.
One of the candidates is a dark photon, which is a gauge boson associated with a hidden U(1) gauge symmetry.
There are various known scenarios to produce dark photon particles or excitations as dark matter:
gravitational production~\cite{Graham:2015rva,Ema:2019yrd,Ahmed:2020fhc,Kolb:2020fwh,Li:2021fao,Sato:2022jya,Redi:2022zkt},
gravitational thermal scattering~\cite{Tang:2017hvq,Garny:2017kha}, 
production through axion-like couplings~\cite{Agrawal:2018vin,Co:2018lka,Bastero-Gil:2018uel,Moroi:2020has}, 
Higgs dynamics~\cite{Dror:2018pdh,Nakayama:2021avl},
kinetic couplings~\cite{Salehian:2020asa,Firouzjahi:2020whk,Nakai:2022dni,Adshead:2023qiw}, 
thermal production with the bose-enhancement effect~\cite{Yin:2023jjj}, 
and cosmic strings~\cite{Long:2019lwl,Kitajima:2022lre}. 
Moreover, a coherent oscillation of a dark photon field can also account for dark matter.

Initially, a minimal setup for misalignment production of dark photon dark matter was originally proposed in Ref.~\cite{Nelson:2011sf}, which does not work due to the Hubble mass term for the canonically normalized gauge field.
To cancel out this Hubble mass term, an introduction of a non-minimal coupling of dark photons to gravity was proposed~\cite{Arias:2012az}, although it is also excluded by the existence of ghost modes~\cite{Nakayama:2019rhg}.
Currently, one of the viable scenarios for the generation of a coherent dark photon dark matter is that proposed in Ref.~\cite{Kitajima:2023fun}.%
\footnote{As another scenario, vector dark matter with a time-dependent mass is studied in Ref.~\cite{Kaneta:2023lki}.}
This scenario is based on the model where the inflaton $\phi$ coupled with the dark photon through the kinetic function as $f^2(\phi) F_{\mu \nu} F^{\mu \nu}$~\cite{Nakayama:2019rhg}.
To avoid the observational constraints pointed out in Ref.~\cite{Nakayama:2020rka}, they introduced a curvaton field to the model in Ref.~\cite{Nakayama:2019rhg}.

In Ref.~\cite{Nakayama:2020rka}, the original model is ruled out due to the combination of the statistical anisotropy of the curvature perturbations and the isocurvature perturbations.
The former arises because the coherent U(1) gauge field determines one specific direction.
In this regard, an SU(2) gauge field can take an isotropic configuration thanks to its three gauge degrees of freedom.
Actually, such a configuration is dynamically realized in some inflationary models such as gauge-flation~\cite{Maleknejad:2011sq,Maleknejad:2011jw,Maleknejad:2012fw} and chromo-natural inflation~\cite{Adshead:2012kp}.
In chromo-natural inflation, the inflaton is coupled to the SU(2) gauge field through a topological coupling, $\phi F_{\mu \nu}^a \tilde{F}^{a \mu \nu}$, and the inflaton velocity assists the isotropic configuration of the SU(2) gauge field (see Ref.~\cite{Komatsu:2022nvu} for a recent review).
Although the original chromo-natural inflation model with a cosine-type potential is excluded due to a significant production of gravitational waves~\cite{Dimastrogiovanni:2012ew, Adshead:2013qp, Adshead:2013nka}, 
this constraint can be evaded by considering a modification of the axion potential~\cite{Maleknejad:2016qjz,Caldwell:2017chz}.%
\footnote{%
We can also consider that the homogeneous gauge field arises after the CMB scale exits the horizon~\cite{Obata:2014loa, Obata:2016tmo, Domcke:2018rvv, Fujita:2022jkc}. 
Even in this case, the constraint from the isocurvature perturbations is absent since the homogeneous gauge field does not have isocurvature perturbations on the CMB scale.
} 
In fact, such a gravitational waves production is suppressed with a low inflation scale~\cite{Komatsu:2022nvu}, which will be our focus (see low scale axion inflation models, such as the multi-natural inflation~\cite{Czerny:2014wza, Czerny:2014xja, Czerny:2014qqa, Higaki:2014sja}, ALP inflation~\cite{Daido:2017wwb, Daido:2017tbr, Takahashi:2019qmh, Takahashi:2021tff, Takahashi:2023vhv}, hybrid QCD axion inflation~\cite{Narita:2023naj}, and the UV completions~\cite{Czerny:2014xja,Czerny:2014qqa,Higaki:2014sja,Murai:2023gkv,Croon:2014dma,Higaki:2015kta, Higaki:2016ydn}).
In this case, the scenario is free from the constraint of isocurvature perturbations~\cite{Adshead:2013nka, Adshead:2016omu}.

In this paper, we use this mechanism to generate coherent vector dark matter while avoiding the constraints from the curvature and isocurvature perturbations.
Our scenario includes two important stages: the dynamics of an axion and SU(2) gauge field during inflation and the oscillation of a homogeneous gauge field after inflation.%
\footnote{
    Coherent oscillation of SU(2) gauge field as a dark matter has been considered in Ref.~\cite{Elahi:2022hgj}, although the model considered there suffers from serious ghost instability.
    The modified kinetic function along the line of Ref.~\cite{Nakayama:2019rhg} may not help the situation since it is not clear whether isotropic gauge field configuration appears or not and hence it is subject to constraints studied in Ref.~\cite{Nakayama:2020rka}.
}
During inflation, an axion drives inflation and sources the homogeneous mode of an SU(2) gauge field through the topological coupling.
After inflation, the axion decays into standard model particles and stops sourcing the SU(2) gauge field.
Then, the SU(2) gauge field independently oscillates due to its self-coupling term.
Whether the SU(2) gauge symmetry is spontaneously broken as the cosmic temperature decreases or initially broken during inflation, the SU(2) gauge field background is inherited by a massive gauge field in the late time universe. 
The field starts to oscillate due to its mass and behaves as non-relativistic matter. 
Then, the massive gauge field can account for dark matter.
Finally, giving a higher dimensional mixing term with standard model photon, we show that the mechanism is consistent for the misalignment production of the dark photon dark matter by taking account of the observational and experimental constraints. 
We also discuss the phenomenological implications such as dark matter search and dark radiation prediction.

The rest of this paper is organized as follows.
In Sec.~\ref{sec: SU(2) during inflation}, we discuss the dynamics of the axion and SU(2) gauge field and the homogeneous configuration of the SU(2) gauge field during inflation.
In Sec.~\ref{sec: post inflation}, we investigate the evolution of the gauge field after inflation and evaluate the abundance of the coherently oscillating vector dark matter. 
As a realization of the mechanism, we study a model in which the SSB is induced by an SU(2) doublet vacuum expectation value. 
In Sec.~\ref{sec: dark photon}, we introduce the mixing term to photon and show that it can apply to the misalignment production of the dark photon dark matter.
Sec.~\ref{sec: Summary} is devoted to the summary and discussion of our results.

\section{Generation of homogeneous SU(2) gauge field during inflation}
\label{sec: SU(2) during inflation}

First, we focus on the dynamics during inflation.
The relevant terms of the Lagrangian are given by
\begin{align}
    \mathcal{L}_\mathrm{inf} 
    =
    \frac{1}{2} \partial_\mu \phi \partial^\mu \phi
    - V(\phi)
    - \frac{1}{4} F_{\mu \nu}^a F^{a \mu \nu}
    + \frac{1}{4 f} \phi F_{\mu \nu}^a \tilde{F}^{a \mu \nu}
    \ ,
\end{align}
where $V(\phi)$ is the potential of the axion field $\phi$, and $f$ controls the strength of the topological coupling.
Although our discussion also works for  $\phi$ being a spectator field as in Ref.~\cite{Dimastrogiovanni:2016fuu}, we consider that $\phi$ is the inflaton as the minimal scenario.
In the following, we do not specify the form of $V(\phi)$.
The field strength and its dual of the SU(2) gauge field are defined by
\begin{align}
    F_{\mu \nu}^a 
    &\equiv
    \partial_\mu A_\nu^a - \partial_\nu A_\mu^a - g \epsilon^{a b c} A_\mu^b A_\nu^c
    \ ,
    \\
    \tilde{F}^{a \mu \nu}
    &\equiv 
    \frac{\epsilon^{\mu \nu \rho \sigma}}{2 \sqrt{-\det[g_{\mu\nu}]}}
    F_{\rho \sigma}^a
    \ ,
\end{align}
where $A_\mu^a$ is the SU(2) gauge field, $g$ is the SU(2) gauge coupling, $g_{\mu \nu}$ is the space-time metric, and $\epsilon^{a b c}$ and $\epsilon^{ \mu \nu \rho \sigma}$ are totally anti-symmetric tensors. Here, we neglect the vector field mass. This is justified if the Hubble parameter (Eq.\,\eqref{Hubble}) during inflation is much higher than the gauge boson mass. As we will see in the next section, this condition is always satisfied by requiring the correct dark matter abundance even if the SU(2) is spontaneously broken before inflation.

In the following, we adopt the temporal gauge, $A_0^a = 0$, and focus on homogeneous modes of the axion and gauge fields: 
\begin{align}
    \phi(t, \bm{x})
    =
    \phi(t)
    \ , \quad 
    A_i^a(t, \bm{x})
    =
    A_i^a(t)
    \ .
\end{align}
We also assume the flat FLRW metric for the space-time metric,
\begin{align}
    \mathrm{d} s^2
    =
    g_{\mu \nu} \mathrm{d} x^\mu \mathrm{d} x^\nu
    =
    \mathrm{d} t^2 - a(t)^2 (\mathrm{d} x^2 + \mathrm{d} y^2 + \mathrm{d} z^2)
    \ .
\end{align}
The time evolution of the scale factor, $a(t)$, is governed by the Friedmann equation, 
\begin{align}
    \label{Hubble}H^2
    \equiv 
    \left( \frac{\dot{a}}{a} \right)^2 
    =
    \frac{\rho_\phi
    + \rho_A}{3 M_\mathrm{Pl}^2}
    \ ,
\end{align}
where the dot represents a derivative with respect to $t$, $M_\mathrm{Pl} \simeq 2.4 \times 10^{18}$\,GeV is the reduced Planck mass, and the energy densities of the inflaton and gauge field are given by
\begin{align}
    \rho_\phi
    &=
    \frac{1}{2} \dot{\phi}^2 
    + V(\phi) 
    \ ,
    \\
    \rho_A 
    &=
    \frac{1}{2 a^2} \dot{A}_i^a \dot{A}_i^a 
    + \frac{g^2}{4 a^4}
    \left[
        \left( A_i^a A_i^a \right)^2 
        - A_i^a A_i^b A_j^a A_j^b
    \right]    
    \ ,
    \label{eq: rho_A with Aia}
\end{align}
respectively.

The equations of motion for the homogeneous fields are given by
\begin{align}
    \ddot{\phi} + 3 H \dot{\phi} + \frac{\partial V}{\partial \phi}
    &=
    - \frac{g}{2 f a^3} \epsilon^{i j k} \epsilon^{a b c}
    \dot{A}_i^a A_j^b A_k^c
    \ ,
    \\
    \ddot{A}_i^a
    + H \dot{A}_i^a
    + \frac{g^2}{a^2} \left( 
        A_j^b A_j^b A_i^a - A_j^a A_i^b A_j^b
    \right)
    &=
    \frac{g}{2 f a} \dot{\phi} \epsilon^{i j k} \epsilon^{a b c} A_j^b A_k^c
    \ .
\end{align}
From the equation of motion for the gauge field, we expect that the axion velocity can source the gauge field.
If the axion field rolls down the potential fast enough, it is known that the homogeneous gauge field has an isotropic attractor solution~\cite{Adshead:2012kp,Maleknejad:2013npa,Wolfson:2020fqz,Wolfson:2021fya}, which has the form of 
\begin{align}
    A_i^a
    =
    \delta_i^a a(t) Q(t)
    \ .
    \label{eq: isotropic Aia}
\end{align}
We substitute this ansatz into the equations of motion and obtain
\begin{align}
    \ddot{\phi} + 3 H \dot{\phi} + \frac{\partial V}{\partial \phi}
    &=
    - \frac{3 g}{f} Q^2 \left( \dot{Q} + H Q \right)
    \ , 
    \\
    \ddot{Q} + 3 H \dot{Q} 
    + \left( 
        \dot{H} + 2 H^2
    \right) Q
    + 2 g^2 Q^3 
    &=
    \frac{g}{f} Q^2 \dot{\phi}
    \ .
\end{align}
The energy density of the gauge field, Eq.~\eqref{eq: rho_A with Aia}, is also simplified as
\begin{align}
    \label{rhoA}\rho_A
    =
    \frac{3}{2} \left( \dot{Q} + H Q \right)^2
    + \frac{3}{2} g^2 Q^4
    \ .
\end{align}

Here, we approximate that the Hubble parameter is constant and focus on the stationary solution, $\ddot{Q}, H\dot{Q} \ll H^2 Q$ and $\ddot{\phi} \ll H \dot{\phi}$, with which the energy density $\rho_A$ is also approximately constant.
Note that we keep $\dot{\phi}$ because the motion of $\phi$ is important in that it sources the gauge field.
Then, we simplify the equations of motion as
\begin{align}
    3 H \dot{\phi} + \frac{\partial V}{\partial \phi}
    &=
    - \frac{3 g}{f} H Q^3
    \label{eq: EoM for phi homogeneous SR}
    \ , 
    \\
    2 H^2 Q
    + 2 g^2 Q^3 
    &=
    \frac{g}{f} Q^2 \dot{\phi}
    \ .
\end{align}
It is convenient to introduce dimensionless parameters that characterize the amplitude of the gauge field, the velocity of the axion field, and the strength of the $\phi F \tilde{F}$ coupling by
\begin{align}
    m_Q
    \equiv 
    \frac{g Q}{H}
    \ , \quad 
    \xi 
    \equiv 
    \frac{\dot{\phi}}{2 f H}
    \ , \quad 
    \Lambda_Q \equiv 
    \frac{Q}{f}
    \ ,
\end{align}
respectively.
Obviously, the equations of motion have a trivial solution, 
\begin{align}
    Q = 0
    \ , \quad 
    \dot{\phi}
    =
    - \frac{1}{3H} \frac{\partial V}{\partial \phi}
     \ .
\end{align}
There is also a nontrivial solution with $Q \neq 0$.
In the case of $\Lambda_Q \gg 1$, the backreaction from the gauge field in the right-hand side of Eq.~\eqref{eq: EoM for phi homogeneous SR} becomes much larger than the Hubble friction term.
Then, we obtain
\begin{align}
    m_Q
    = 
    \left(
        - \frac{g^2 f}{3 H^4} \frac{\partial V}{\partial \phi} 
    \right)^{1/3}
    \ ,
    \label{eq: mQ solution}
    \\
    m_Q^3 - \xi m_Q^2 + m_Q
    =
    0
    \label{eq: EoM for m_Q}
    \ ,
\end{align}
which leads to 
\begin{align}
    m_Q = 0, \frac{\xi \pm \sqrt{\xi^2 - 4}}{2}
    \ .
\end{align}

\begin{figure}[t]
    \centering
    \includegraphics[width=.8\textwidth ]{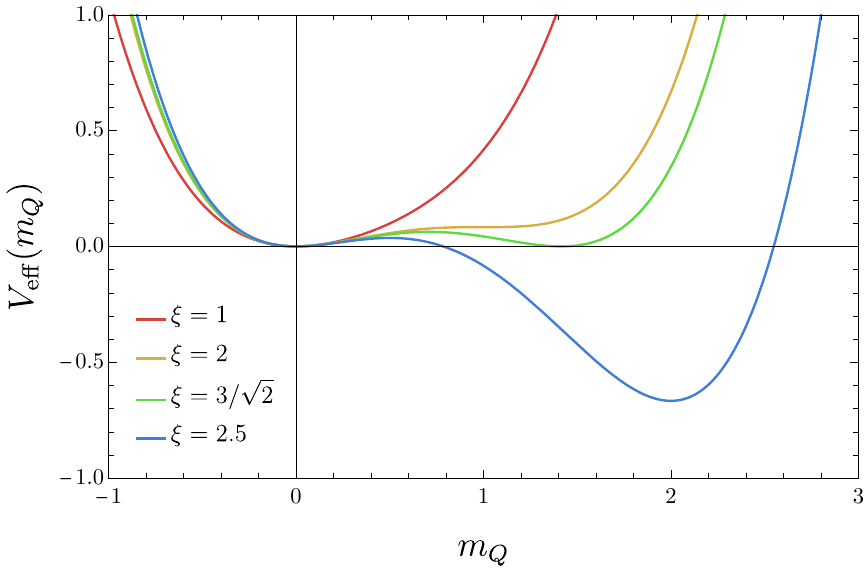}
    \caption{%
        Effective potential for $\xi$ = $1$, $2$, $3/\sqrt{2}$, $2.5$.
        The potential has a local minimum at $m_Q \neq 0$ for $\xi > 2$, and it becomes the global minimum for $\xi > 3/\sqrt{2}$. 
    }
    \label{fig: Potential}
\end{figure}

To see which solution realizes, we define the effective potential for $m_Q$ by integrating the left-hand side of Eq.~\eqref{eq: EoM for m_Q} with respect to $m_Q$ as
\begin{align}
    V_\mathrm{eff}(m_Q)
    \equiv 
    \frac{1}{4} m_Q^4 - \frac{\xi}{3} m_Q^3 + \frac{1}{2} m_Q^2
    \ .
\end{align}
We show the shape of the effective potential in Fig.~\ref{fig: Potential}.
Regardless of the value of $\xi$, this potential has a local minimum at $m_Q = 0$.
In addition, it has another minimum at $m_Q = (\xi + \sqrt{\xi^2 - 4})/2$ for $\xi > 2$, which becomes the global minimum for $\xi > \xi_\mathrm{c} \equiv 3/\sqrt{2} \simeq 2.12$.
Note that the other solution of Eq.~\eqref{eq: EoM for m_Q}, $m_Q = (\xi - \sqrt{\xi^2 - 4})/2$, corresponds to the local maximum of $V_\mathrm{eff}$ for $\xi > 2$.
Thus, we expect that the nontrivial solution is realized for $\xi \gtrsim \xi_\mathrm{c}$.

When the gradient of the axion potential becomes large enough, the slow-roll condition is violated and inflation ends.
The slow-roll parameter is given by
\begin{align}
    \epsilon_H
    \equiv 
    -\frac{\dot{H}}{H^2}
    =
    \epsilon_\phi + \epsilon_E + \epsilon_B
    \ ,
\end{align}
where
\begin{align}
    \epsilon_\phi 
    \equiv 
    \frac{\dot{\phi}^2}{2 M_\mathrm{Pl}^2 H^2}
    \ , \quad 
    \epsilon_E 
    \equiv 
    \frac{\left( \dot{Q} + H Q \right)^2}{M_\mathrm{Pl}^2 H^2}
    \ , \quad 
    \epsilon_B
    \equiv 
    \frac{g^2 Q^4}{M_\mathrm{Pl}^2 H^2}
    \ .
\end{align}
With the stationary solution, in particular, the slow-roll parameters are given by
\begin{align}
    \epsilon_\phi 
    =
    \frac{2 \xi^2 f^2}{M_\mathrm{Pl}^2}
    \ , \quad 
    \epsilon_E 
    = 
    \frac{m_Q^2 H^2}{g^2 M_\mathrm{Pl}^2}
    \ , \quad 
    \epsilon_B
    = 
    \frac{m_Q^4 H^2}{g^2 M_\mathrm{Pl}^2}
    \ .
\end{align}
For $\Lambda_Q \gg 1$, we obtain
\begin{align}
    \frac{\epsilon_B}{\epsilon_\phi}
    =
    \frac{m_Q^2 \Lambda_Q^2}{2 \xi^2}
    \gg 
    1
    \ ,
\end{align}
and thus $\epsilon_B$ is dominant in the total slow-roll parameter $\epsilon_H$.
Thus, we define the end of inflation by $\epsilon_B = 1$ in the following.

\section{Vector dark matter from axion-SU(2) inflation}
\label{sec: post inflation}

In this section, we study the evolution of vector dark matter in our mechanism by using a concrete example.
\begin{figure}[t]
    \centering
    \includegraphics[width=.8\textwidth ]{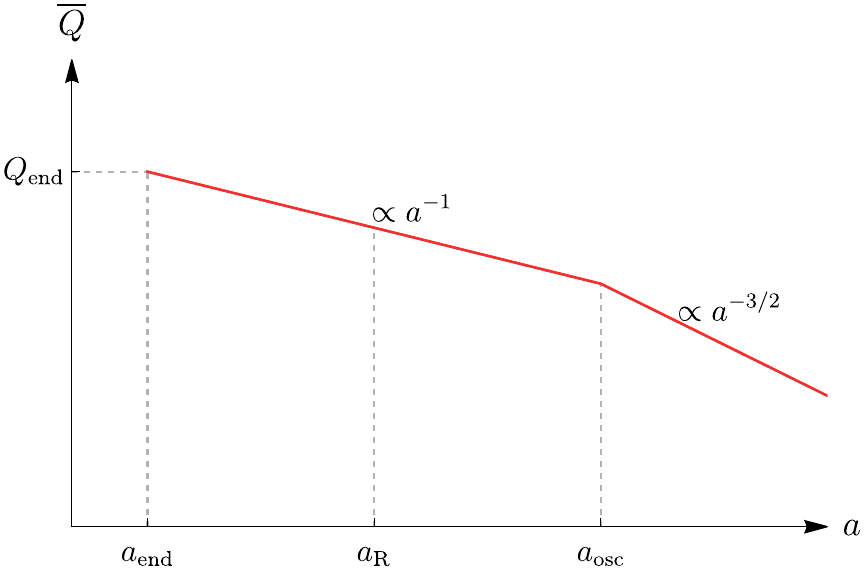}
    \caption{%
        Time evolution of the amplitude of the gauge field, $\overline{Q}$ in the pre-inflationary scenario.
        At the end of inflation, the SU(2) gauge field has a nonzero background, $Q_\mathrm{end}$, due to the axion-gauge field dynamics during inflation.
        Then, the gauge field starts to oscillate due to the quartic potential of $Q$ originating from the self-coupling and $\overline{Q}$ decreases as $\propto a^{-1}$.
        Once the mass potential induced by the SSB overcomes the quartic potential, the gauge field comes to behave as non-relativistic matter with $\overline{Q}\propto a^{-3/2}$.
    }
    \label{fig: Q evolution}
\end{figure}
We consider the simplest scenario with an SU(2) doublet, $\Phi$, to give mass to the gauge fields. 
The relevant Lagrangian is given by
\begin{align}
    \mathcal{L}_\mathrm{SSB}
    =
    D_\mu \Phi^\dagger D^\mu \Phi
    - V_\mathrm{d}(\Phi)
    \ ,
\end{align}
where the covariant derivative, $D_\mu$, is defined by
\begin{align}
    D_\mu \Phi 
    =
    \partial_\mu \Phi 
    - i g A_\mu^a \frac{\sigma^a}{2} \Phi
    \ .
\end{align}
Here, $\sigma^a$ is the Pauli matrix, and the potential is assumed to be
\begin{align}
    V_\mathrm{d}(\Phi)
    =
    \frac{\lambda}{4} 
    \left(
        \Phi^\dagger \Phi - v^2
    \right)^2
    \ ,
\end{align}
with a positive constant $\lambda$. 
Considering that this potential has a negative mass term, $- \lambda v^2 \Phi^\dagger \Phi/2$, $\Phi$ is expected to acquire a nonzero expectation value either before or after inflation. 

Once the SU(2) symmetry is spontaneously broken, the gauge field acquires a mass term given by
\begin{align}
    \mathcal{L} 
    \supset 
    \frac{g^2 v^2}{4} A_\mu^a A^{a \mu}
    \ ,
\end{align}
which corresponds to a mass of $m = g v/\sqrt{2}$.
In this case, the three components of the SU(2) gauge field become massive at the SSB. 
The masses are the same. This is a specific spectrum with the single double Higgs field breaking the SU(2).

In the following, we will discuss the scenario with the SSB before inflation (pre-inflationary scenario) and after inflation (post-inflationary scenario).
In Fig.~\ref{fig: Q evolution}, we summarize the time evolution of the amplitude of the oscillating gauge field, $\overline{Q}$ in the pre-inflationary scenario.

\subsection{Pre-inflationary scenario and abundance of the massive vector fields}
\label{subsubsec: abundance for doublet}

Let us first consider the possibility that the SU(2) is always broken during and after inflation. 
This is the case if the Higgs boson mass satisfies 
\begin{align}
    m_\Phi \gg H 
    \ ,
    \label{mFggH}
\end{align}
during inflation, and the matter effect never lets the symmetry recover. 
Here $m_\Phi\simeq \sqrt{\lambda} v$ at the tree-level. 
Although the gauge fields contribute to the effective mass of the Higgs field, it is of the order of $g Q = m_Q H$ and is not important due to the condition \eqref{mFggH}. 
Again, we consider that the gauge boson mass satisfies $m\ll H$ around the end of inflation. This will be shown to be a self-consistent condition in the following. 

We also emphasize that the discussion in this section is within an effective theory and it is not sensitive to UV completion other than the mass spectrum of the three vector bosons.
After inflation, the inflaton typically starts to oscillate around its potential minimum,
and the gauge field deviates from the static solution given by Eq.~\eqref{eq: mQ solution}. 

Although the evolution of the axion and gauge field depends on the details of reheating, we assume that the gauge field decouples from the inflaton during reheating.
As shown in Appendix~\ref{app: axion-gauge reheating}, this assumption can be justified by setting the mass and coupling of the inflaton appropriately.
There, we also discuss the condition for the inflaton not to decay to the gauge field immediately (for instance, see also the model in Ref.\,\cite{Moroi:2020has}).

First, we evaluate the amplitude of the gauge field at the end of inflation.
From the condition of $\epsilon_B = 1$, we obtain the gauge field amplitude at the end of inflation as
\begin{align}
   \label{mQgQ} m_{Q,\mathrm{end}}
    =
    \frac{g Q_\mathrm{end}}{H_\mathrm{end}}
    \simeq 
    \sqrt{\frac{g M_\mathrm{Pl}}{H_\mathrm{end}}}
    \ ,
\end{align}
where the subscript ``end'' denotes quantities at the end of inflation.
Note that this relation leads to $Q_\mathrm{end} = M_\mathrm{Pl}/m_{Q,\mathrm{end}}$, and the gauge field amplitude is smaller than the Planck scale by $m_{Q,\mathrm{end}}$.
To satisfy $\Lambda_Q = Q/f \gg 1$ at the end of inflation, we require $f \ll M_\mathrm{Pl}$ in the following.
Similarly, we will use the subscripts ``R'', ``SSB'', and ``osc'' for quantities at the completion of reheating, the SSB of SU(2) (in the next subsection), and the onset of gauge field oscillations due to the mass, respectively.
While the nontrivial solution exists for $m_Q \geq 1$, the backreaction from the enhanced perturbations becomes non-negligible for large $m_Q$ depending on the value of $g$~\cite{Fujita:2017jwq, Ishiwata:2021yne}. 
The requirement becomes $m_Q\lesssim O(10)$ for $g=10^{-O(10)}$, which we will focus on in the following.
In the following, we will treat $m_{Q,\mathrm{end}}$ as a fixed parameter, which relates $H_\mathrm{end}$ and $g$ as
\begin{align}
    H_\mathrm{end}
    =
    \frac{g M_\mathrm{Pl}}{m_{Q,\mathrm{end}}^2}
    \ .
\end{align}

Even after inflation, there is no reason for the gauge field to deviate from the isotropic configuration in Eq.~\eqref{eq: isotropic Aia}.
Thus, we follow the evolution of $Q$ in the following.
Once the gauge field decouples from the axion, the equation of motion for $Q$ becomes
\begin{align}
    \ddot{Q} + 3 H \dot{Q} 
    + \left( 
        \dot{H} + 2 H^2
        + m^2
    \right) Q
    + 2 g^2 Q^3 
    &=
    0
    \ .
    \label{eq: EoM for Q after inf}
\end{align}
Until the completion of reheating, the universe is dominated by the oscillating inflaton, which is assumed to behave as non-relativistic matter as usual.
Thus, $H = 2/(3t)$ and we obtain
\begin{align}
    \ddot{Q} + 3 H \dot{Q} 
    + \left( \frac{H^2}{2} + m^2 \right) Q
    + 2 g^2 Q^3 
    &=
    0
    \ .
\end{align}
This equation of motion can be understood as that for a scalar field with the mass including the Hubble-induced term and the quartic potential.
Just after inflation, $m_{Q,\mathrm{end}} > 1$, and the dynamics of the gauge field is controlled by the quartic potential since $m$ is smaller than $H_\mathrm{end}$.
Then, $Q$ oscillates with a decreasing amplitude of $\overline{Q} \propto a^{-1}$.
Since $H$ decreases as $\propto t^{-1} \propto a^{-3/2}$, $m_Q$ remains larger than unity even after inflation.
Thus, we consider that the amplitude of $Q$ decreases as $a^{-1}$ until the completion of reheating.

The completion of reheating is represented by  
\begin{align}
    \Gamma_\phi = H
    =
    \sqrt{\frac{\pi^2 g_{*,\mathrm{R}}}{90}}\frac{T_\mathrm{R}^2 }{M_\mathrm{Pl}}
    \ ,
\end{align}
where $\Gamma_\phi$ is the decay width%
\footnote{We include the matter effect in the decay rate of the inflaton, and $\Gamma_{\phi}$ may also depend on the temperature. 
In our discussion with generic $V(\phi)$, $f$ is a free parameter for having the SU(2) stationary solution during inflation, and we focus on the possibility that the decay rate of $\phi$ into the dark vector bosons is suppressed. 
}
of the inflaton, $T_\mathrm{R}$ is the reheating temperature, and $g_*$ is the relativistic degrees of freedom for the energy density.
Since the Hubble parameter evolves as $H \propto a^{-3/2}$ during reheating, the ratio of the scale factors between the end of inflation and the completion of reheating becomes 
\begin{align}
    c
    \equiv
    \frac{a_\mathrm{R}}{a_\mathrm{end}}
    =
    \left( 
        \frac{H_\mathrm{end}}{\Gamma_\phi}
    \right)^{2/3}
    \ .
\end{align}
With the factor $c \geq 1$, the reheating temperature is represented as
\begin{align}
    T_\mathrm{R}
    =
    c^{-3/4}
    \left( \frac{90}{\pi^2 g_{*,\mathrm{R}}} \right)^{1/4}
    \frac{\sqrt{g} M_\mathrm{Pl}}{m_{Q,\mathrm{end}}}
    \equiv 
    c^{-3/4} T_\mathrm{max}
    \ ,
    \label{TR_Tmax}
\end{align}
where $T_\mathrm{max}$ denotes the maximum value of $T_\mathrm{R}$ realized in the case of instantaneous reheating.

After the completion of reheating, the universe is dominated by radiation.
During the radiation dominated era, $H = 1/(2t)$, the equation of motion for $Q$ becomes
\begin{align}
    \ddot{Q} + 3 H \dot{Q} 
    + m^2 Q
    + 2 g^2 Q^3 
    &=
    0
    \ .
\end{align}
As long as $m \lesssim g \overline{Q}$, $Q$ continues to oscillate with an amplitude $\propto a^{-1}$.
Thus, the amplitude of $Q$ after inflation is roughly given by
\begin{align}
    \overline{Q} 
    \simeq 
    \sqrt{\frac{M_\mathrm{Pl} H_\mathrm{end}}{g}}
    \frac{a_\mathrm{end}}{a}
    \ .
\end{align}
Then, it comes to follow $a^{-3/2}$ once the mass term becomes dominant when
\begin{align}
    m
    \simeq 
    g \overline{Q}
    \ .
\end{align}
From this condition, we obtain the scale factor at the onset of oscillation by the mass potential, $a_\mathrm{osc}$, as 
\begin{align}
    \frac{a_\mathrm{end}}{a_\mathrm{osc}}
    \simeq 
    \frac{m_{Q,\mathrm{end}} m}{g M_\mathrm{Pl}}
    \ .
\end{align}
The corresponding temperature is given by
\begin{align}
    T_\mathrm{osc}
    &\simeq 
    \left( 
        \frac{g_{*s,\mathrm{R}} a_\mathrm{R}^3}{g_{*s,\mathrm{osc}} a_\mathrm{osc}^3}
    \right)^{1/3}
    T_\mathrm{R}
    \nonumber \\
    &\simeq 
    \left( 
        \frac{g_{*s,\mathrm{R}}}{g_{*s,\mathrm{osc}}}
    \right)^{1/3}
    \frac{m_{Q,\mathrm{end}} m}{g M_\mathrm{Pl}}
    c T_\mathrm{R}
    \ ,
    \label{Tosc}
\end{align}
where $g_{*s}$ is the relativistic degrees of freedom for the entropy density.
After that, the amplitude of $Q$ is given by
\begin{align}
    \overline{Q} 
    \simeq 
    \frac{m}{g}
    \left( 
        \frac{a_\mathrm{osc}}{a}
    \right)^{3/2}
    \ .
\end{align}
Thus, the amplitude at the matter-radiation equality becomes
\begin{align}
    \overline{Q}_\mathrm{eq}
    &\simeq 
    \frac{m}{g}
    \left( 
        \frac{a_\mathrm{osc}}{a_\mathrm{eq}}
    \right)^{3/2}
    \nonumber \\
    &\simeq 
    \frac{1}{c^{3/8} g^{1/4}}
    \left( \frac{\pi^2 g_{*,\mathrm{R}}}{90} \right)^{3/8}
    \sqrt{\frac{g_{*s,\mathrm{eq}}} {g_{*s,\mathrm{R}}}}
    \frac{T_\mathrm{eq}^{3/2}}{m^{1/2}}
    \ .
\end{align}
We get the energy density of the oscillating gauge field as
\begin{align}
    \rho_{Q,\mathrm{eq}}
    &=
    \frac{3}{2} m^2 \overline{Q}_\mathrm{eq}^2
    \nonumber \\
    &\simeq 
    \frac{3}{2 c^{3/4} g^{1/2}}
    \left( \frac{\pi^2 g_{*,\mathrm{R}}}{90} \right)^{3/4}
    \frac{g_{*s,\mathrm{eq}}} {g_{*s,\mathrm{R}}}
    m T_\mathrm{eq}^3
    \ .
\end{align}
As a result, the current density parameter for the dark vector boson is given by
\begin{align}
    \Omega_{Q,0}
    &=
    \Omega_{\mathrm{m},0} \frac{\rho_{Q,\mathrm{eq}}}{\rho_\mathrm{r,eq}}
    \nonumber \\
    &\simeq 
    \Omega_{\mathrm{m},0}
    \frac{1}{2 c^{3/4} g^{1/2}}
    \frac{g_{*,\mathrm{R}}}{g_{*s,\mathrm{R}}}
    \frac{g_{*s,\mathrm{eq}}}{g_{*,\mathrm{eq}}}
    \left( \frac{\pi^2 g_{*,\mathrm{R}}}{90} \right)^{-1/4}
    \frac{m}{T_\mathrm{eq}}
    \ .
    \label{eq: Omega Q formula}
\end{align}
This is 
\begin{align}
    \Omega_{Q,0} h^2
    =
    0.1
    \left( \frac{m}{1\,\mathrm{meV}} \right)
    \left( \frac{g}{10^{-11}} \right)^{-1/2}
    \left( \frac{c}{10^3} \right)^{-3/4}
    \ ,
\end{align}
where $h$ is the reduced Hubble constant.%
\footnote{%
    Note that $\Omega_{Q,0} \sim \Omega_{\rm m,0}(T_{\rm osc}/cT_{\rm eq})$, i.e., the dark matter abundance is determined by $T_{\rm osc}$ and $c$ for $m < g \overline{Q}_\mathrm{SSB}$ (see Eqs.~(\ref{TR_Tmax}) and~(\ref{Tosc})). 
    Interestingly, therefore, when the dark matter abundance is fixed, $T_{\rm osc}$ only depends on the period of the reheating phase.
    Since $T_{\rm osc}$ controls the redshift that the dark matter is formed from radiation and affects the small-scale structure, if this scenario is correct, precisely measuring small-scale structure can be a probe of the reheating phase.
}
Here, we assumed $g_{*,\mathrm{R}} = g_{*s,\mathrm{R}} = 106.75$, $g_{*s,\mathrm{eq}} = 3.36$, $g_{*,\mathrm{eq}} = 3.91$ and used $\Omega_{\mathrm{m},0} h^2= 0.1424$ and $T_\mathrm{eq} = 3387 T_0$~\cite{Planck:2018vyg} with $T_0 = 2.755$\,K being the current CMB temperature~\cite{Fixsen:2009ug}.

So far, we have assumed $m\ll H_{\rm end}$. 
This is consistent with the abundance estimation. 
One notes from the condition of $\epsilon_B = 1$, $\rho_Q\sim g^2 Q^4\sim H_{\rm end}^2 M_{\rm Pl}^2$ soon after inflation. 
If $m \gtrsim H_{\rm end}$ were satisfied, the dark matter would dominate over the universe. 

We also note the need for a non-vanishing period of the reheating phase, $c\gg 1$, because otherwise, this scenario explaining the abundance would predict the dark matter too hot.
This is because the comparable energy density soon after inflation as thermal bath would explain the dark matter with $T_{\rm osc}\sim $ eV, similar to the hot dark matter scenario. 
It is similarly excluded because the free-streaming length is too long. 
The detailed constraint on the free-streaming will be discussed in the last subsection. 

\subsection{Post-inflationary scenario}

Next, let us consider the scenario that the symmetry is spontaneously broken after inflation. 
This is the case with Eq.\,\eqref{mFggH} not satisfied or the matter effect after inflation lets the symmetry restore.%
\footnote{In this case, the abundance estimation does not change due to the conservation of the adiabatic invariant.}
The stochastic effect of the dark Higgs field is not important. 
This is because the dark Higgs acquires $Q$-induced mass $\sim g Q \sim m_{Q,\rm end} H_{\rm end}$, which is parametrically larger than the Hubble parameter since $m_{Q,\rm end}$ is parametrically larger than one.
This traps the dark Higgs field in the symmetric phase~\cite{Kitajima:2021bjq,Nakagawa:2022knn} unless a delicate cancellation of the mass terms occurs.

To discuss the scenario let us, for simplicity, consider that the $\Phi$ sector is in thermal equilibrium with the visible sector after inflation. 
Thus, they share the same cosmic temperature, $T$.
The thermal mass usually takes over the $Q$-induced mass unless the relevant coupling is extremely small. 
Thus, in the discussion of the cosmology after inflation, we neglect the $Q$-induced mass. 
If $\Phi$ is thermalized with the temperature $T$, it acquires an effective mass of $\sim \sqrt{\lambda} T$, which leads to $T_\mathrm{SSB} \sim v$. 
If $\Phi$ is coupled to the standard model Higgs, $H_{\rm SM}$, through $\lambda_{\Phi H_{\rm SM}} |\Phi|^2 |{H_{\rm SM}}|^2$, the effective mass of $\sim \sqrt{\lambda_{\Phi H_{\rm SM}}} T$ is induced at high temperatures and we obtain $T_\mathrm{SSB} \sim \sqrt{\lambda/\lambda_{\Phi H_{\rm SM}}} v$.
In the following, we parameterize the SSB temperature by $T_\mathrm{SSB} \equiv \sqrt{\Lambda} v$.
Using $m$, the temperature at the SSB, $T_\mathrm{SSB}$, can be written as
\begin{align}
    T_\mathrm{SSB}
    =
    \frac{\sqrt{2 \Lambda} m}{g}
    \ .
\end{align}

After the SSB, $Q$ behaves like a scalar field with quadratic and quartic potential terms.
The evolution of such a field depends on which term is dominant in the potential at this moment.
Let us first  consider the case  that the quartic term is dominant at the SSB:
\begin{align}
    m 
    <
    g \overline{Q}_\mathrm{SSB}
    =
    \frac{g M_\mathrm{Pl}}{m_{Q,\mathrm{end}}} 
    \frac{a_\mathrm{end}}{a_\mathrm{SSB}}
    \ ,
    \label{eq: mass vs quartic at SSB}
\end{align}
which leads to
\begin{align}
    g 
    <
    \sqrt{\frac{\pi^2 g_{*,\mathrm{R}}}{90 c}} 
    \left( \frac{g_{*s, \mathrm{SSB}}}{g_{*s, \mathrm{R}}} \right)^{2/3}
    2 \Lambda 
    \ .
\end{align}
When the quartic potential is dominant, the amplitude of $Q$ after inflation decreases as $a^{-1}$ until the mass becomes important. 
Then, the abundance estimation is the same as the pre-inflationary case. 

On the other hand, if the quadratic potential is dominant at the SSB, we obtain $T_\mathrm{osc} = T_\mathrm{SSB}$.
Here, we assume that the SSB occurs with a timescale longer than that of the oscillations of $Q$.
This is true, for example, if the SSB occurs in a few Hubble time scale (note that $g \overline{Q} \gg H$ is satisfied much after inflation before the SSB since $g \overline{Q} \propto a^{-1}, H\propto a^{-3/2}$ or $a^{-2}$, and $g\overline{Q}_{\rm end}\sim H_{\rm end}$.)
In this case, due to the conservation of the adiabatic invariant, the amplitude of $Q$ just after the SSB, $\overline{Q}_\mathrm{aft}$, is written using that just before the SSB, $\overline{Q}_\mathrm{bef}$, as
\begin{align}
    m \overline{Q}_\mathrm{aft}^2
    \simeq 
    g \overline{Q}_\mathrm{bef}^3
    \ .
\end{align}
Since the adiabatic invariant evolves as $a^{-3}$ both before and after the SSB, $m \overline{Q}^2$ after the SSB does not depend on when the SSB occurs.
In particular, $m \overline{Q}^2$ or $\overline{Q}$ after the SSB takes the same value for $m < g \overline{Q}_\mathrm{SSB}$ and $m > g \overline{Q}_\mathrm{SSB}$, which can also be checked through a direct calculation.
As a result, the abundance estimation in the post-inflationary case is the same as the pre-inflationary case although the time evolution of $\bar{Q}$ is different from that in Fig.~\ref{fig: Q evolution} if $m > g \overline{Q}_\mathrm{SSB}$.

\subsection{Phenomenological aspects of the scenario}
\label{subsubsec: constraints for doublet}

\begin{figure}[t]
    \centering
    \includegraphics[width=.45\textwidth ]{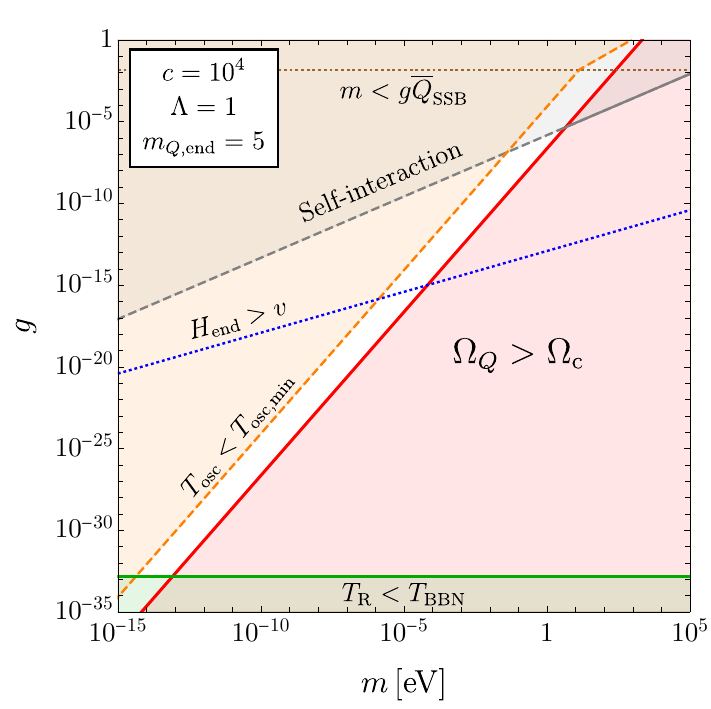}
    \hspace{5mm}
    \includegraphics[width=.45\textwidth ]{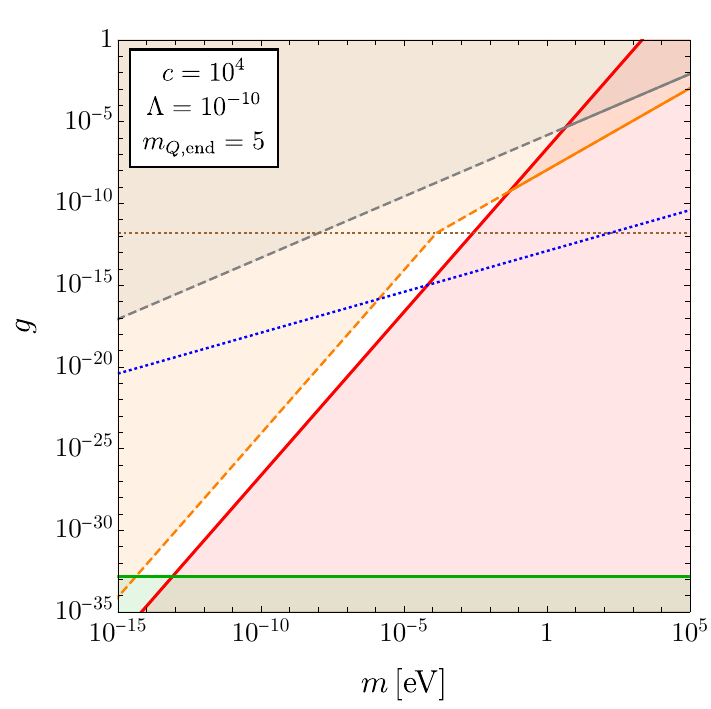}
    \caption{%
        Abundance of vector dark matter for $c = 10^4$ and $m_{Q,\mathrm{end}} = 5$.
        We adopt $\Lambda = 1$ and $10^{-10}$ in the left and right panels, respectively.
        The dark vector accounts for the observed abundance of dark matter on the red line, and the red-shaded region is excluded by over-abundance.
        The gray-shaded region is excluded by the constraint on self-interaction of dark matter if the dark vector accounts for all dark matter.
        The orange-shaded region is excluded for all dark matter because the formation of dark matter is too late.
        Above the red line, the gray and orange lines are shown by dashed lines because the dark vector field accounts for a part of dark matter.
        The green-shaded region is excluded because the reheating temperature is lower than the BBN temperature.
        The blue dotted line shows the typical threshold for the pre/post-inflationary scenario.
        Above the blue dotted line, the SU(2) symmetry is expected to be restored during inflation though the exact threshold depends on $\lambda$ and the thermal history of the dark sector.
        On the horizontal brown dotted line, the quadratic and quartic potential is comparable at the SSB in the post-inflationary scenario.
    }
    \label{fig: DM doublet}
\end{figure}

Here, we discuss some constraints and the reach of future observations for this scenario. 
By including the constraints that we will discuss, the parameter region where the dark vector field can account for dark matter is shown in Fig.~\ref{fig: DM doublet}.
In this simple setup for breaking the SU(2), one obtains the vector boson dark matter in the mass range of $m=10^{-13}\,\text{--}\,1\,$eV. 
As we will discuss in the last section, the range can be widened in a more generic setup.

\paragraph{Self-interaction}
First, we consider the self-interaction of dark matter, which is constrained by the observations of galaxy clusters through, e.g., dissociative mergers~\cite{Markevitch:2003at,Randall:2008ppe}, strong lensing~\cite{Meneghetti:2000gm}, oscillations of the brightest cluster galaxy~\cite{Kim:2016ujt,Harvey:2018uwf}, and the density profile~\cite{Eckert:2022qia}.
The constraint is typically given by
\begin{align}
    \frac{\sigma}{m}
    \lesssim
    \mathcal{O}(0.1\,\text{--}\,1)\,\mathrm{cm^2/g}
    \ ,
\end{align}
where $\sigma$ is the self-scattering cross section.
If the three components of the SU(2) gauge field constitute dark matter,%
\footnote{If the mass degeneracy is broken by some mechanism or higher dimensional terms, the heavier components may decay, and the constraint from the self-interaction disappears. 
In addition, the realistic bound may be weaker because of the wave-like feature. The dark matter wave collision may be more likely to happen in a forward direction due to the Bose enhancement effect.}
its self-interaction is estimated as
\begin{align}
    \frac{\sigma}{m}
    \sim
    \frac{g^4}{m^3}
\end{align}
from the dimension analysis. 
Then, the conservative constraint, $\sigma/m \lesssim 1\,\mathrm{cm^2/g}$, is translated into a constraint on $g$ and $m$ as 
\begin{align}
    m \gtrsim 60\,\mathrm{MeV} \times g^{4/3}
    \ .
\end{align}

\paragraph{Big-bang nucleosynthesis}
The reheating temperature should be high enough for the successful big bang nucleosynthesis (BBN).
Here, we require
\begin{align}
    T_\mathrm{R}
    >
  T_{\rm BBN} \simeq  10\,\mathrm{MeV}
    \ ,
\end{align}
which leads to
\begin{align}
    g
    &>
    c^{3/2}
    \left( \frac{\pi^2 g_{*,\mathrm{R}}}{90} \right)^{1/2}
    \frac{m_{Q,\mathrm{end}}^2}{M_\mathrm{Pl}^2}
    (10\,\mathrm{MeV})^2
    \nonumber \\
    &\simeq 
    1.4 \times 10^{-39} c^{3/2}
    \ .
\end{align}

\paragraph{Small-scale structures}
In this scenario, the hidden gauge fields at first behave as radiation%
\footnote{%
In this paper, we do not take into account the self-resonance effect due to the $Q$ oscillation in the quartic potential, which should be an interesting topic. If this effect is important, a lot of $Q$ particles may be produced to behave as radiation as well. Thus, it does not change our conclusions. However, the constraints from the small-scale structure can be slightly different.}
and, when the mass potential becomes important at $m \simeq g\overline{Q}$, they come to behave as non-relativistic matter.
If dark matter experiences such a transition from dark radiation, it suppresses the matter perturbations on small scales.
From the observations of small-scale structures, the redshift of the transition, $z_\mathrm{T}$, is bounded from below~\cite{Sarkar:2014bca,Corasaniti:2016epp,Das:2020nwc}.
In particular, the abundance of the Milky Way satellite galaxies gives~\cite{Das:2020nwc}
\begin{align}
    z_\mathrm{T} 
    >
    5.5 \times 10^6
    \ ,
\end{align}
which corresponds to a lower bound of $T_\mathrm{osc}$:
\begin{align}
    T_\mathrm{osc}
    >
    T_\mathrm{osc,min}
    \simeq 
    1.3 \times 10^{-6}\,\mathrm{GeV}
    \ .
\end{align}
As mentioned around Eq.~\eqref{eq: Omega Q formula}, $\Omega_{Q,0}$ can be written as
\begin{align}
    \Omega_{Q,0}
    &\simeq 
    \Omega_{\mathrm{m},0}
    \frac{g_{*,\mathrm{R}}}{g_{*s,\mathrm{R}}}
    \frac{g_{*s,\mathrm{eq}}}{g_{*,\mathrm{eq}}}
    \left( 
        \frac{g_{*s,\mathrm{osc}}}{g_{*s,\mathrm{R}}}
    \right)^{1/3}
    \frac{T_\mathrm{osc}}{2 c T_\mathrm{eq}}
    \ .
\end{align}
By fixing the dark matter abundance as $\Omega_{Q,0}h^2 = \Omega_{\mathrm{c},0}h^2 = 0.11933$~\cite{Planck:2018vyg}, where $\Omega_{\mathrm{c},0}$ denotes the present energy fraction of cold dark matter, we obtain the lower bound on $c$ as
\begin{align}
    c
    >
    c_\mathrm{min}
    &\equiv 
    \frac{\Omega_{\mathrm{m},0}h^2}{\Omega_{\mathrm{c},0}h^2}
    \frac{g_{*,\mathrm{R}}}{g_{*s,\mathrm{R}}}
    \frac{g_{*s,\mathrm{eq}}}{g_{*,\mathrm{eq}}}
    \left( 
        \frac{g_{*s,\mathrm{osc}}}{g_{*s,\mathrm{R}}}
    \right)^{1/3}
    \frac{T_\mathrm{osc,min}}{2T_\mathrm{eq}}
    \nonumber \\
    &\simeq
    1.1 \times 10^3
    \ ,
\end{align}
in the pre-inflationary scenario or the post-inflationary scenario with $m < g\overline{Q}_\mathrm{SSB}$.
In the post-inflationary scenario with $m > g\overline{Q}_\mathrm{SSB}$, the formation of dark matter is delayed until the SSB and the constraint from $z_\mathrm{T}$ becomes more severe.

\paragraph{Fate of $\Phi$}
We also consider the fate of the doublet, $\Phi$.
In the pre-inflationary scenario, $\Phi$ is decoupled from the visible sector, and the energy density of the dark sector apart from the homogeneous gauge field is negligible.
The fate of $\Phi$ is unimportant unless it is produced thermally after inflation, which is discussed soon.

On the other hand, in the post-inflationary scenario, $\Phi$ is assumed to be thermalized with the visible sector.
One possibility is through a sizable Higgs portal interaction. Assuming the $\Phi$ mass is smaller than the weak scale, which is consistent with the parameter region shown in Fig.\,\ref{fig: DM doublet}, the $\Phi$ sector is thermalized until the electroweak phase transition if $\lambda_{\Phi H_{\rm SM}}^2 T > H$ at $T\sim 100\,$GeV. This leads to $\lambda_{\Phi H_{\rm SM}}\gtrsim 10^{-8}.$ 
 We note that the transverse modes of the gauge fields rarely thermalize because of the small gauge couplings in the parameter region of interest.

Soon after the electroweak phase transition, one finds that the $\Phi$ fields naturally decouple from the visible sector because they only interact with the standard model particles via higher dimensional terms.
In this case, the four degrees of freedom of $\Phi$ decouples from the visible sector soon after the electroweak phase transition.
If the quartic coupling is not very small, the entropy of the dark sector is transferred into the three degrees of freedom of the longitudinal modes of the hidden gauge fields when the temperature of the dark sector after the SSB  becomes smaller than the dark Higgs mass $\sqrt{\lambda} v$.
The produced hidden gauge fields behave as a relativistic component.
We can estimate their contribution to dark radiation as
\begin{align}
    \Delta N_\mathrm{eff}
    \simeq 
    \frac{3 \times \left( \frac{10.75}{106.75} \times \frac{4}{3} \right)^{4/3}}{2 \times 7/8}
    \simeq 
    0.12
    \ ,
\end{align}
which will be within reach of future CMB observations~\cite{CMB-S4:2016ple,CMB-HD:2022bsz}.%
\footnote{%
    Here, we assume that the production of the gauge field excitations after inflation is negligible since we are interested in small $g$ (see Fig.~\ref{fig: DM doublet}).
    Moreover, the production during inflation is also negligible due to the suppression by $\mathcal{O}(H_\mathrm{inf}^2/M_\mathrm{Pl}^2)$ with $H_\mathrm{inf}$ being the Hubble parameter during inflation.
}
We note that the momenta of the longitudinal vector bosons contributing to $\Delta N_{\rm eff}$ are much higher than that of the coherent oscillation, and they coexist in the later Universe.

In addition, the portal coupling can be probed from the invisible decay of the standard model-like Higgs boson. 
In fact, when the dark Higgs boson is also light, future lepton colliders are quite powerful for probing the invisible decay of both Higgs bosons~\cite{Haghighat:2022qyh}.

\section{Dark photon dark matter with kinetic mixing} 
\label{sec: dark photon}
\begin{figure}[t]
    \centering
    \includegraphics[width=.8\textwidth]{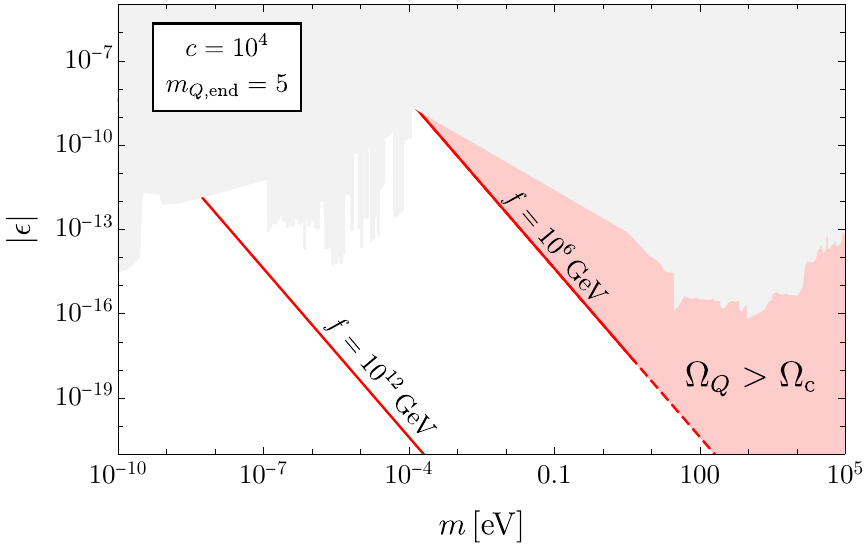}
    \caption{%
        Kinetic mixing of the dark vector and photon.
        The gray-shaded region is constrained by the observations and experiments~\cite{Caputo:2021eaa}.
        Here, the constraints from the dark matter searches are shifted by $\sqrt{3}$ since only 1/3 of the dark matter mixes with the photon in this scenario.
        The red lines correspond to all dark matter, and the constrained region for $\Lambda = 1$ is shown in the dashed lines (see Fig.~\ref{fig: DM doublet}). 
        In the red-shaded region, the vector dark matter is overproduced.
    }
    \label{fig: Kinetic mixing doublet}
\end{figure}

So far, we have discussed the SU(2) vector boson as the dominant dark matter produced from the misalignment mechanisms by giving a mass from the vacuum expectation value of the SU(2) doublet. One of the vector bosons can be identified as the dark/hidden photon since
the SU(2) gauge field can mix with a standard model photon through kinetic mixing by introducing a higher dimensional term:
\begin{align}
    \mathcal{L}
    \supset 
    \frac{\kappa}{2f^2}\Phi^\dagger F^a_{\mu\nu} \sigma^a F^{\gamma}_{\mu\nu} \Phi
    \ ,
\end{align}
where $\kappa$ is a dimensionless coupling constant, and $F^\gamma$ is the field strength tensor of the standard model photon.
Here, we assumed that this term is suppressed by $f$ in analogy to the $\phi F \tilde{F}$ term.
When $\Phi$ acquires a nonzero expectation value, this term leads to the kinetic mixing as 
\begin{align}
    \mathcal{L}
    \supset 
    \frac{\kappa m^2}{g^2 f^2}
    F^\gamma_{\mu \nu} F^{\mu \nu 3}
    \ ,
\end{align}
where we assumed $\Phi = (v, 0)^\mathrm{T}$ without loss of generality.
This term leads to the kinetic mixing between the photon and one component of the SU(2) gauge field:
\begin{align}
    \mathcal{L}
    \supset 
    -\frac{\epsilon}{2}F_{\mu \nu}^\gamma F^{\mu \nu 3}
    \ , \quad 
    \epsilon 
    \equiv 
    -\frac{2 \kappa m^2}{g^2 f^2}
    \ .
\end{align}
We show the constraint from the kinetic mixing for $\kappa = 1$ in Fig.~\ref{fig: Kinetic mixing doublet}.
The red lines correspond to the parameter region where the gauge field explains all dark matter, and the constrained regions are shown in the dashed lines (see Fig.~\ref{fig: DM doublet}).
The kinetic mixing scales as $f^{-2}$ and can be probed by future experiments and observations depending on the value of $f$.
Interestingly, the predicted kinetic mixing in the region of the post-inflationary scenario may be within reach of future experiments searching for dark photon absorption by various materials~\cite{Hochberg:2016ajh,Hochberg:2016sqx,Bloch:2016sjj,Hochberg:2017wce,Knapen:2017ekk,Chigusa:2020gfs,Chigusa:2021mci,Mitridate:2021ctr,Chen:2022pyd,Mitridate:2022tnv,Chigusa:2023hmz}. This can be also probed in LAMPOST~\cite{Baryakhtar:2018doz},~BREAD~\cite{BREAD:2021tpx} (Dish-antenna experiment \cite{Horns:2012jf}), MADMAX~\cite{Gelmini:2020kcu}, ALPHA~\cite{Gelmini:2020kcu}, and Dark E-field~\cite{Godfrey:2021tvs}.

We note that only 1/3 of the dark matter mixes with the photon. Since the sensitivity of the dark matter direct detection experiment is $\propto \epsilon^2 \Omega_{\rm \gamma'} \approx \epsilon^2 \Omega_{\rm Q,0}/3 $, we need to recast the bound/sensitivity reach of $\epsilon$ by a factor of $\sqrt{3}$. 
This is not the case for the astrophysics bound or tabletop experiment in which the dark photon particles are produced irrelevant to the axion-SU(2) dynamics. 
In other words, the difference in the sensitivity by a factor of $\sqrt{3}$ between the dark matter search, such as the haloscopes mentioned above, and tabletop experiments, such as light shining through the wall experiments~\cite{Abel:2006qt,Arias:2010bh,Bahre:2013ywa, Ortiz:2020tgs}, can also distinguish our scenario from the other dark photon scenarios. 

\section{Summary and discussion}
\label{sec: Summary}

We have discussed a new mechanism to generate a coherently oscillating dark vector field from the axion-SU(2) gauge field dynamics during inflation. 
This mechanism can evade the bounds from isocurvature perturbation, statistical anisotropy, and the ghost modes that exist in the usual misalignment production of the dark photon.
    
In the scenarios we have proposed, all of the SU(2) gauge fields behave as coherent oscillating dark matter while only one component, i.e., 1/3 of the total dark matter, can couple to the standard model particle via mixing.
The resulting dark matter mass range can be much lighter than the eV scale and can be probed in various haloscopes as well as helioscopes and tabletop experiments. 
The difference by $1/\sqrt{3}$ in the mixing parameter between the searches as dark matter and other searches is a unique prediction of our scenario. 
In particular, the coherent oscillating vector boson that mixes with a photon has a polarization towards a fixed direction over a very large spatial scale before the structure formation. 
If this polarization does not change after the structure formation,
the scenario predicts daily/annual modulation of the dark matter signals~\cite{Caputo:2021eaa}.
Since the regions that can be probed from the dark photon absorption by various materials~\cite{Hochberg:2016ajh,Hochberg:2016sqx,Bloch:2016sjj,Hochberg:2017wce,Knapen:2017ekk,Chigusa:2020gfs,Chigusa:2021mci,Mitridate:2021ctr,Chen:2022pyd,Mitridate:2022tnv,Chigusa:2023hmz} and in the experiments, Super-CDMS~\cite{Bloch:2016sjj}, LAMPOST~\cite{Baryakhtar:2018doz},~BREAD~\cite{BREAD:2021tpx} (Dish-antenna experiment \cite{Horns:2012jf}), MADMAX~\cite{Gelmini:2020kcu}, ALPHA~\cite{Gelmini:2020kcu}, Dark E-field~\cite{Godfrey:2021tvs} and light shining through the wall experiments~\cite{Abel:2006qt,Arias:2010bh,Bahre:2013ywa, Ortiz:2020tgs} is in the range of the post-inflation scenario (see Fig.\ref{fig: DM doublet}),
the discovery of the hidden photon in them, together with the dark radiation of $\Delta N_{\rm eff}=\mathcal{O}(0.1)$ is another smoking-gun signal of the scenario.

In fact, the early production of the SU(2) gauge field during inflation can induce gravitational waves, and the $Q^4$ oscillation would have self-resonance, which also contributes to the gravitational waves. 
These effects, however, do not change our conclusions because the inflation scale is low. 

So far, we have discussed a model for the realization of the mechanism by introducing an SU(2) doublet Higgs to break the SU(2) gauge group spontaneously. 
Our mechanism works in a more generic setup. 
For instance, we can break the SU(2) by introducing Higgs triplet(s) to give masses for the gauge boson dark matter, which leads to a more generic mass difference.%
\footnote{When the mass is different, the pre-inflationary scenario may have an additional constraint from the statistical anisotropy.}
In this case, we have monopoles, and the polarization of the dark photon dark matter is more non-trivial. This will be discussed elsewhere. 
In addition, our mechanism should even work with generic SU($N$) with $N>2$, which is also believed to have isotropic attractor solutions of homogeneous gauge field~\cite{Fujita:2021eue,Fujita:2022fff,Murata:2022qzz}.
Alternatively, we can even break the SU(2) in a non-linear sigma model without introducing a Higgs field. 

More generically, we can reduce the lower mass range for dark matter. This is the case if we have the gauge coupling changes effectively. The amplitude and the energy density of the vector field scales with $\overline{Q} \propto g^{-1/3}$ and $g^2 \overline{Q}^4\propto g^{2/3}$ by noting the adiabatic invariant $g \overline{Q}^3.$ Reducing $g$ makes the comoving energy density smaller and $m$ for the transition, say $z_\mathrm{T}\sim 10^7$, smaller since the amplitude is larger.
As a result, we can have the vector dark matter mass as small as the fuzzy dark matter one. 
The decrease of $g$ may be naturally realized from a symmetry breaking of SU(2)$\times$ SU(2)$\to$ SU(2) or from a field excursion of a moduli field relevant to the gauge coupling.

\begin{acknowledgments}
We would like to thank Fuminobu Takahashi for useful discussions in the early stages of this project.
This work was supported by JSPS KAKENHI Grant Nos. 18K13537 (T.F.),
20H05851 (W.Y.), 20H05854 (T.F.), 21K20364 (W.Y.), 22H01215 (W.Y.), 22K14029 (W.Y.), and 23KJ0088 (K.M.).
\end{acknowledgments}

\appendix 
\section{Axion and gauge field during reheating}
\label{app: axion-gauge reheating}

In Sec.~\ref{sec: post inflation}, we assumed that reheating lasts for the period parameterized by $c$ and that the gauge field decouples from the axion during reheating (see Eq.~\eqref{eq: EoM for Q after inf}).
In this appendix, we discuss the validity of these assumptions. 

First, we investigate the condition for the inflaton not to immediately decay to the gauge field.
In the following, we assume that the inflaton potential after inflation is given by a quadratic form with a mass of $m_\phi$, while we do not specify the shape of the inflaton potential during inflation.
Through the coupling of $\phi F_{\mu \nu}^a \tilde{F}^{a \mu \nu}/(4f)$, the oscillating inflaton decays to a pair of the gauge boson particles with a decay rate of 
\begin{align}
    \Gamma_{\phi \to AA}
    =
    \frac{3 m_\phi^3}{64 \pi f^2}
    \ .
\end{align}
Thus, we require
\begin{align}
    \frac{\Gamma_{\phi \to AA}}{H_\mathrm{end}}
    \ll 
    c^{-3/2}
    \ ,
\end{align}
where we used $H \propto a^{-3/2}$ during reheating.

Moreover, the inflaton decay can be enhanced due to the stimulated emission of the vector bosons.
Including the stimulated emission, the distribution function of the gauge boson with a comoving momentum $k$ can be given as~\cite{Moroi:2020has}
\begin{align}
  f_k (t\rightarrow\infty) =
   \frac{1}{2}
  \left( e^{2\bar{f}} - 1 \right)
  \theta \left( \frac{m_\phi}{2} - k \right)
  \ ,
  \label{f_k(exp)}
\end{align}
where $\theta$ is the Heaviside step function, and $\bar{f}$ is given by 
\begin{align}
\label{expo}
    \bar{f} 
    &\equiv
    \frac{32\pi^2 \Gamma_{\phi \to AA} n_\phi}{H m_\phi^3}
    \nonumber \\
    &\simeq 
    \frac{9 \pi}{2}
    \frac{H M_\mathrm{Pl}^2}{m_\phi f^2}
    \ ,
\end{align}
where $n_\phi = \rho_\phi/m_\phi$ is the number density of $\phi$ with $\rho_\phi \simeq 3 M_\mathrm{Pl}^2 H^2$ being the energy density of $\phi$.
Here, $\bar{f}$ takes the maximum value $\bar{f}_\mathrm{end}$ just after the end of inflation.
To ensure that the inflaton does not decay into the dark vector bosons through the stimulated emission, we also require
\begin{align}
    r_\mathrm{se}
    \equiv 
    \frac{m_\phi^3}{32 \pi^2 n_\phi} 
    \frac{e^{2\bar{f}_\mathrm{end}} - 1}{2}
    \ll
    1
    \ .
\end{align}
Note that $r_\mathrm{se}$ becomes $\Gamma_{\phi \to AA}/H$ in the limit of $\bar{f} \ll 1$.

Next, we discuss the dynamics of the axion and gauge field during reheating.
In particular, we numerically simulate the field evolution and discuss the effects of the coupling between the axion and gauge field.
To this end, we write down the equations of motion for $\phi$ and $Q$ as
\begin{align}
    \frac{\mathrm{d}^2 \varphi}{\mathrm{d} x^2} 
    + 3 h \frac{\mathrm{d} \varphi}{\mathrm{d} x} 
    + \varphi
    &=
    - 3 \frac{m_\phi^2}{g^2 f^2} q^2
    \left( \frac{\mathrm{d} q}{\mathrm{d} x} + h q \right)
    \ , 
    \label{eq: EoM for varphi}
    \\
    \frac{\mathrm{d}^2 q}{\mathrm{d} x^2} 
    + 3 h \frac{\mathrm{d} q}{\mathrm{d} x} 
    + \left( 
        \frac{\mathrm{d} h}{\mathrm{d} x} + 2 h^2
    \right) q
    + 2 q^3 
    &=
    q^2 \frac{\mathrm{d} \varphi}{\mathrm{d} x}
    \ ,
    \label{eq: EoM for q}
\end{align}
using the dimensionless quantities
\begin{align}
    x \equiv m_\phi t 
    \ , \quad 
    h \equiv \frac{H}{m_\phi}
    \ , \quad 
    \varphi \equiv \frac{\phi}{f}
    \ , \quad 
    q \equiv \frac{g Q}{m_\phi}
    \ .
\end{align}
Here, we neglected the decay of $\varphi$, which provides a conservative check for the validity of the assumption to neglect the coupling.
Since we assume the matter domination during reheating, the Hubble parameter is given by 
\begin{align}
    h = \frac{2}{3 x} \ .
\end{align}
Using the analytical solution at the end of inflation:
\begin{equation}
\begin{gathered}
    \xi 
    \equiv 
    \frac{1}{2 h} \frac{\mathrm{d} \varphi}{\mathrm{d} x}
    =
    m_{Q,\mathrm{end}} + \frac{1}{m_{Q,\mathrm{end}}}
    \ , 
    \\
    \sqrt{\epsilon_B}
    \equiv 
    \frac{m_{Q,\mathrm{end}}^2 h m_\phi}{g M_\mathrm{Pl}}
    = 1
    \ , \quad  
    \frac{3}{2} M_\mathrm{Pl}^2 h^2
    =
    \frac{1}{2} f^2 \varphi^2
    \ ,
\end{gathered}
\end{equation}
we obtain the initial conditions as
\begin{equation}
\begin{gathered}
    h_\mathrm{i}
    =
    \frac{g}{m_{Q,\mathrm{end}}^2} \frac{M_\mathrm{Pl}}{m_\phi}
    \ , \quad 
    x_\mathrm{i}
    =
    \frac{2}{3 h_\mathrm{i}}
    \ , \quad
    \\
    q_\mathrm{i}
    =
    m_{Q,\mathrm{end}} h_\mathrm{i}
    \ , \quad 
    \left. \frac{\mathrm{d} q}{\mathrm{d} x} \right|_\mathrm{i}
    =
    0
    \ ,
    \\
    \varphi_\mathrm{i}
    =
    \sqrt{3} h_\mathrm{i} \frac{M_\mathrm{Pl}}{f}
    \ , \quad 
    \left. \frac{\mathrm{d} \varphi}{\mathrm{d} x} \right|_\mathrm{i}
    =
    2 h_\mathrm{i} \left( m_{Q,\mathrm{end}} + \frac{1}{m_{Q,\mathrm{end}}} \right)
    \ .
\end{gathered}
\end{equation}
Here, the subscript $\mathrm{i}$ denotes the initial conditions.

\begin{figure}[t]
    \centering
    \includegraphics[width=.8\textwidth ]{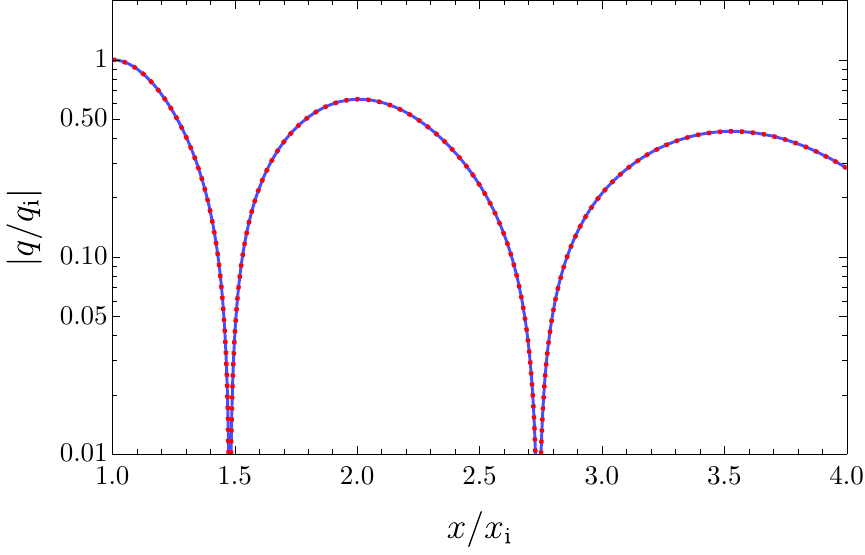}
    \caption{%
        Time evolution of $q$ for $m_\phi = 10^{12}$\,GeV and $f/M_\mathrm{Pl} = 0.004$.
        The blue-solid and red-dotted lines show the results with and without the axion-gauge field coupling, respectively.
    }
    \label{fig: numerical q}
\end{figure}

In the following, we fix $m_{Q,\mathrm{end}} = 5$ as in the main text and take $g = 10^{-12}$ as a typical value.
We consider $m_\phi = 5 \times 10^{10}$, $5 \times 10^{11}$, $10^{12}$\,GeV and $f/M_\mathrm{Pl} = 0.002$, $0.004$, $0.04$.
Every set of these $m_\phi$ and $f$ gives $\Lambda_Q > 1$, $\Gamma_{\phi \to AA}/H_\mathrm{end} < 1$, and $r_\mathrm{se} < 1$.
We show the time evolution of $q$ in Fig.~\ref{fig: numerical q}.
Here, we choose $m_\phi = 5 \times 10^{10}$\,GeV and $f/M_\mathrm{Pl} = 0.004$, which gives $\Lambda_Q = 50$, $\Gamma_{\phi \to AA}/H_\mathrm{end} = 2.1 \times 10^{-7}$, and $r_\mathrm{se} = 1.8 \times 10^{-6}$.
For comparison, we also show the time evolution of $q$ without the coupling between the axion and gauge field (i.e., neglecting the right-hand side of Eqs.~\eqref{eq: EoM for varphi} and \eqref{eq: EoM for q}).
We see that the coupling has a negligible effect on the evolution of $q$.
We obtained similar results for the other choices of $m_\phi$ and $f$.
Thus, by appropriately taking the parameters, $m_\phi$ and $f$, we can safely neglect the coupling term in discussing the evolution of $q$ during reheating.

\bibliographystyle{JHEP}
\bibliography{Ref}

\end{document}